\documentclass[useAMS,usenatbib]{mn2e}
\usepackage{epsfig}
\usepackage{amsmath}
\usepackage{multirow}
\title[]{
Effects of a planetesimal debris disk on stability scenarios
for the extrasolar planetary system HR 8799
}
\author[]{
Alexander Moore \& Alice C. Quillen															\\
Department of Physics and Astronomy, University of Rochester, Rochester, NY 14627, USA  	\\
}

\begin{document}
\maketitle

\begin{abstract}
HR 8799 is a four planet system that also hosts a debris disk. By numerically integrating
both planets and a planetesimal disk, we find interactions between an exterior planetesimal
disk and the planets can influence the lifetime of the system. We first consider resonant
planetary configurations that remained stable for at least 7 Myrs sans debris disk. An
exterior debris disk with only $\sim1\%$ the mass of the outermost planet (approximately a
Neptune mass) was sufficiently large enough to pull the system out of resonance after 2 to
6 Myrs. Secondly, we consider configurations which are unstable in less than a few hundred
thousand years. We find that these can be stabilized by a debris disk with a mass of more
than $\sim10\%$ that of the outermost planet. Our two sets of simulations suggest that
estimates of the long term stability of a planetary system should take into account the
role of the debris disk.
\end{abstract}

\section{Introduction}
Photometric surveys conducted with Kepler space telescope \citep{borucki11} in addition to
radial velocity surveys such as those conducted by \citet{wright09} have indicated that
multiple planet systems are common. Numerical integrations can be used to determine if these
systems are stable and estimate the time to first orbit crossing event or collision
\citep{gozdziewski09,reidemeister09,fabrycky10,marshall10,marois10}. Any configuration that
has a short lifetime is considered less likely. Consequently, integrations can be used to
place constraints on both the orbital elements and masses of the planets. These numerical
investigations often neglect planetesimals. However, extrasolar planetary systems can harbor
planetesimal debris disks \citep{su09,moro-martin10}.

HR 8799 is a $1.5 \pm 0.3 M_{sol}$ A5V star found $39.4 \pm 1.0 pc$ from Earth
\citep{marois08}. It has at least four planets which have been directly imaged
\citep{marois10}. Mass estimates for the planets, even taking into account HR 8799's young
age of $60^{+100}_{-30}$ Myr determined by various techniques \citep{marois08}, or
$30^{+20}_{-10}$ as part of the Columba association \citep{marois10}, have values of
$10 \pm 3$ $M_{Jupiter}$ for HR 8799c, d, and e and $7^{+4}_{-2}$ $M_{J}$ for HR 8799b for an
assumption of a 60 Myr age, and masses of $7^{+3}_{-2}$ $M_{J}$ for HR 8799c, d, e and
$5 M_{J}$ for an assumption of a 30 Myr age. Projected separations for the planets of HR 8799e, d, c
and b from the star are observed to be 14.5, 24, 38 and 68 AU, respectively \citep{marois10}.
However, this measurement lacks a long baseline helpful for constraining the planets'
positions. Astrometric measurements with a much longer baseline taken with \textit{Hubble Space
Telescope} in 1998 for the outer three planets HR 8799b, c and d confirm reported values for
HR 8799b and add new observations for HR 8799c and d \citep{soummer11}.

Dynamical studies of HR 8799 indicate that it is likely in a 4:2:1 dual mean motion
resonance (MMR). In other words, the inner two planets are in a 2:1 mean motion resonance while
the outer pair are also in a 2:1 mean motion resonance. This architecture is required to
explain how the system has remained stable over its observed age
\citep{gozdziewski09,fabrycky10,reidemeister09,marshall10,marois10}.
Possible orbital configurations determined by numerical simulations are summarized by
\citet{moro-martin10}.

HR 8799 also has an inner debris disk ranging from 6-15 AU, an outer debris disk which is
thought to extend from 90 to 300 AU and a dusty halo out to $\sim1000$ AU \citep{su09}.
The specific details of this model will be discussed further in subsection 1.2.

Planetesimal debris disks can influence the long term stability of a planetary system.
Within the context of the `Nice' model \citep{tsiganis05}, planets migrate due to
interactions with planetesimals, and instability occurs when two planets cross a strong
mean motion resonance. \citet{thommes08} considered systems put in resonance by a gas disk,
a possible scenario explaining HR 8799's current configuration. However, after the gas disk
had been depleted, they found that planetary interactions with the remnant planetesimal disk
tended to remove these systems from resonances and induce dynamical instability.

It is in this context that we examine the role of the debris disk in affecting the stability
of the HR 8799 system. First, we investigate if it is possible to delay the onset of
instability for an initially highly unstable orbital configuration. Then we consider a
configuration with a long lifetime and determine whether a debris disk can cause instability.

\subsection{Resonant Structure of HR 8799}
As discussed in the introduction, initial solutions tested by
\citet{gozdziewski09,reidemeister09,fabrycky10} had suggested that HR 8799b, c and d were most
likely in 4:2:1 dual mean motion resonances. Nearly all orbital configurations that remain
stable during a reverse integration for the estimated age of the system which also agree with
the observed orbital elements and nominal masses for the planets minimally called for HR 8799c
and d to be in a 2:1 MMR. Furthermore, the dual mean motion 4:2:1 resonance
followed by the inner pair of HR 8799c and d in a 2:1 MMR allow for the largest possible range
of masses that still produce simulations stable over the lifetime of the system. Alternate
solutions included a 2:1 MMR among the outer pair HR 8799b and c as well as a few finely tuned
solutions. In a later analysis, \citet{marshall10} found that when the three planet configuration
is placed in the 4:2:1 MMR with low planet eccentricities, HR 8799 could survive the observed age
of the system and potentially longer. Re-reduction of Hubble space telescope astrometric data of
HR 8799 by \citet{soummer11}, originally taken in 1998, robustly supports the previous 4d:2c:1b
dual MMR hypothesis with results which have only small departures from the exact integer period
ratios in previously published data.

This previous dynamical work precedes the discovery of the innermost planet HR 8799e by \citet{marois10}.
However, simulations by \citet{marois10} have indicated that the extra planet only places more
restrictions on the possible system architectures. In order for those simulations to remain
stable over the minimum estimated lifetime of the HR 8799 system, the planets e, d and c are
required to be in a 4:2:1 MMR along with masses on the minimum end of the estimated values.
Note that the 4:2:1 dual mean motion resonance is still the most likely configuration for this
system to be in if long term stability is desired - the difference being which planets 
are in the mean motion resonance, their projected masses due to dynamical considerations, and
the maximum age of the system which remains stable given these orbital elements.

\subsection{Distribution and Total Mass of HR 8799's Debris Disk}
The total mass and distribution of the debris found in the disk affects the dynamics, migration rate,
extent of migration, and the smoothness of migration. However, the total mass of the inner and outer
debris disk along with the dust halo is not well known for HR 8799. Estimating the total mass in
debris within a disk is difficult to compute accurately for both observational and theoretical reasons.
Solid (non-gaseous) matter, which emits continuously in the $\mu m$ to mm wavelengths produces nearly
all of the radiation \citep{beckwith96}. However, while the total cross section of the dust particles
is many orders of magnitude larger than that of larger objects (m to km sized asteroids, comets, and
planetesimals), it is these later objects which comprise a majority of the total mass of the debris.
These objects are not bright enough to individually detect in the wavelengths that they emit. But
estimating the total mass of the debris is important to determining the dynamics.

The halo and disk dust has been modelled by \citet{su09}. The halo mass is estimated to be $1.9 \times
10^{-2} M_{\oplus}$ with a radius of up to 1000 AU. Estimating the dust mass of the other disk components
is more difficult - particularly for the cold outer disk because the inner and outer edges are not well known.
In the case of the inner edge, it was first thought to be at 90 AU from temperature estimates while the
outer radius had been modelled at 300 AU to account for all the observational constraints, giving a total
dust mass of $1.2 \times 10^{-1} M_{\oplus}$. The inner disk is quite warm at $\sim150 K$, allowing for a
very low dust mass estimate of $1.1 \times 10^{-6} M_{\oplus}$. A brief summary of the modelled parameters
that were used including assumed surface densities, inner and outer radii, minimum and maximum grain size,
as well as total mass can be found in \citet{su09} in their Table 3.

Recent submillimeter observations have added some additional constraints. \citet{hughes11} examined
HR 8799 and its debris disk with the Submillimeter Array at $880 \mu m$ - a wavelength which is suitable
for examining larger dust grains. Low signal-to-noise prevented a full multi-parameter modelling of the
dust at this wavelength but a combination of the SMA data and spectral energy distributions (SEDs) seem
to rule out a narrow ring of dust and favor a broad outer ring starting with an inner edge at $\sim 150$ AU. 

Scaling up from these modelled dust masses to a total debris disk mass can be difficult. Small dust
grains are typically evacuated in debris disks by a combination of Poynting-Robertson drag and radiation
blow-out (depending on grain size) on very short timescales relative to the age of an average debris
disk. To constantly replenish the dust in a system that is many millions of years old, a collisional
cascade is required \citep{safronov69,dohnanyi69,williams94,tanaka96,kenyon99,kenyon01,kenyon02}. 

It is assumed that the differential number distribution of the particles in mass is a power law which
takes the form
\begin{equation}
dn(m) = Am^{-\alpha}dm
\end{equation}
or simply
\begin{equation}
dn(a) = Ca^{2-3\alpha}da
\end{equation}
in radius. The steady-state solution depends on an $\alpha$ index of $\frac{11}{6}$. This parameter
has been estimated empirically and numerically in the previously mentioned literature and, assuming
that $\alpha$ is determined only by the mass-dependence of the collision rate and that the model is
self-similar, can be shown to be that value analytically \citep{tanaka96}.

To estimate the total mass of the disk, we integrate the differential number distribution
times the mass of these objects at a constant density from the minimum to maximum sizes in the cascade
\begin{equation}
M_{T} = \int _{a_{min}} ^{a_{max}} M(a) dn(a).
\end{equation}
The constant in our number distribution can be set by using the model of the outer debris disk by
\citet{su09}. This model includes the minimum and maximum grain sizes that were used in order to create a
synthetic SED which was matched to the observed data. With the dust mass estimate from that model, and a minimum
and maximum grain size of $10\mu m$ and $1000\mu m$ respectively, the constant $C$ can be computed.

To find the total mass of the disk we repeat the previous calculation only with a new $a_{max}$ that
corresponds to the largest radius objects in our collisional cascade. The total mass that this integrand
will yield is entirely set by the upper bound because there are many orders of magnitude between the
minimum and maximum object sizes. Unfortunately, determining $a_{max}$ is difficult.

Assuming the cascade operates for the system age \citet{quillen07} estimated $a_{max}$, assuming an alpha
parameter of 11/6, that used only observable properties of the debris disk. As per their equation 16,
\begin{equation}
\begin{split}
a_{top}& \approx 5.4 \rm{km} \left(\frac{\lambda}{10 \mu m}\right) \left(\frac{M_{\star}}{M_{\odot}}\right)^{8/3}
\left(\frac{r}{100 \rm{AU}}\right)^{-14/3}\\
& \quad \times \left(\frac{Q ^{\star} _{D}}{2 \times 10^6 \rm{erg} \cdot \rm{g^{-1}}}\right)^{-5/3}
\left(\frac{t_{age}}{10^7 \rm{yr}} \right)^2 \left(\frac{h}{0.02}\right)^{10/3}\\
& \quad \times \left[\frac{\bar{\tau}(\lambda)}{10^{-2}}\right]^2 \left(\frac{f_{\tau}}{4}\right)^{-2}
\end{split}				
\end{equation}
where $h$, the disk aspect ratio, and $\bar{\tau}$, the normal disk opacity at wavelength $\lambda$, are
our disk observables. Other parameters include $M_{\star}$, $\lambda$, r, $Q ^{\star} _{D}$, $t_{age}$,
and $f_{\tau}$, which correspond to the stellar mass, observation wavelength, the radii at which there
is a break in the surface brightness profile, specific energy, age, and an uncertainty factor.

However, HR 8799 has no constraints on the scale height because the disk is not resolved. We therefore adopt
the $\beta$ Pictoris estimate for $h$ by \citet{quillen07}. $\beta$ Pictoris is also an A star which has a
similar age and mass to HR 8799 as well as an extended debris disk. Estimates for the dust mass around
$\beta$ Pictoris have been found to be $7.8M_{Moon}$, or about $0.096M_{\oplus}$ \citep{holland98}. This is
similar to the $0.12M_{\oplus}$ suggested by \citet{su09} for HR 8799. \citet{su09} also notes that the amount
of excess emission in the HR 8799 disk is similar to that of other debris disks around A stars (See
\citet{su06}'s study of the evolution of debris disks around A stars.) See table 1 by \citet{quillen07} and
references therein for mass, age, and other parameters for $\beta$ Pictoris.

We note that our biggest uncertainty is therefore in $h$ in the above formula, which goes as $\frac{10}{3}$.
This makes even small errors in estimates of $h$ have a large impact on $a_{top}$. However, the goal is not
to determine the exact mass of the disk, only to determine what a reasonable range is.

To find $a_{top}$ we must also compute the opacity at a specific wavelength. The opacity can be measured by
modifying our total dust mass integral. The opacity at a specific wavelength is a measure of the fractional
area covered by particles of radius $a$ (i.e. the opacity depends on the number of particles per unit area
times the cross sectional area). In this way we can relate the number of particles of a specific radius to
the opacity and total disk surface area. Secondly, it is possible to equate the differential number distribution
to $N_{a}$ by
\begin{equation}
\frac{dN(a)}{dln(a)} \equiv N(a) ,
\end{equation}
\citep{quillen07}. Thirdly, we make use of a relation which describes how the opacity scales with the
radius of the object,
\begin{equation}
\tau_{(a)} = \tau_{d} \left( \frac{a}{a_d} \right) ^{3-q}
\end{equation}
where $q = 3.5$ is equivalent to $\alpha = \frac{11}{6}$ from above and $a_d$ and $\tau _d$ are the radius
and opacity at a specific radius/wavelength \citep{quillen07}. Substituting the above three relationships
into our mass computation yields an integral that is dependent only on the radius of the dust. Using
$1000 \mu m$ dust particles as the largest grains we find that the opacity at the specific dust particle radius is
\begin{equation}
\tau _d \sim \frac{M_d}{A}\frac{1}{\rho a_d}
\end{equation}
where $M_d$ is the total mass of the dust, $A$ is the total disk area, $\rho$ is the density of the dust grains, and
$a_d$ is the radius of the largest dust grain in the model. Using dust masses from \citet{su09} as well as a
disk area computed from the inner and outer edges and a maximum grain size from the same, along with a density
appropriate for dust grains $\rho = 1.5-3.0 \frac{g}{cm^3}$, we compute a value for the opacity of 
$\tau_d \sim 10^{-4}$.

With these computed disk observables along with the other known parameters for HR 87999 we find an $a_{top}$
value of $a_{top} \sim 1 km$. This $a_{top}$ yields a disk mass estimate of $M_{disk} \sim 150 M_{\oplus}$ or about
one half of a Jupiter mass. Large debris disks are often attributed masses in the region of $50-100M_{\oplus}$.

While this estimate would place the disk mass usable in our simulations at of order a Jupiter mass or less,
there are a number of caveats to our estimates. We note that our estimate for $a_{top}$ is low compared to
other disks by \citet{quillen07} which place the radius at values anywhere from a few to a few hundred times
larger. Using an estimate for $a_{top}$ in line with those for $\beta$ Pictoris and similar disks would yield
a significantly more massive disk. Also, we recognize
that the masses of the planets are extreme, with four planets of ten Jupiter masses each. It is not inconceivable that
the disk may be significantly more massive than those observed previously. We also note that planetesimals
with a radius of $a_{top}$ are the largest
objects that contribute to the collisional cascade. Significant mass could be found in other, more massive objects
that would have collision times too long to contribute to dust production. Finally, we note that the major phenomenon
discussed in the paper occur at masses of one Jupiter mass or less in the debris disk. It is not required for us
to include much of the more massive disks, however, given the difficulty of estimating the mass and the general
peculiarity and size of HR 8799, we have included fairly generous debris disk masses.

A second more generic estimate of the largest sized objects that can be grown was found by \citet{cuzzi93}.
Using a numerical model which uses the Reynold's averaged Navier-Stokes equations with both turbulence
and full viscosity, they found that it was possible to create 10-100km sized planetesimals. This would
place our disk mass estimate of order $\sim M_J$.

Finally, we can use our own solar system as a reference point. The `Nice' model required approximately 30-50
$M_{\oplus}$ to explain the outward migration and eventual configuration of our own system \citep{tsiganis05}.
This corresponds to a few tenths of a Jupiter mass of debris.

Total debris disk masses used in our simulations varied, but had measurable effects at a Neptune mass, 
or about 17 $M_{\oplus}$. This puts those simulated disk masses in a similar range to that of the `Nice' model.
Larger masses than the `Nice' model are justified by way of the discussion above. Given the uncertainties
in measuring $a_{top}$, it is very difficult to determine if the largest total masses can be used
in our simulations.

The total particle number used in our simulations, 1024, is also comparable to the 1000-5000 particles
used in the `Nice' model simulations. Above we noted that the diameter of objects at the top end of the
collisional distribution is anywhere between one to several hundred km. An $a_{top}$ value similar to
$\beta$ Pictoris of 180km is about 6.5 times smaller than the diameter of Pluto. $a_{top}$ values near
1 km would be approximately one thousand times smaller. This suggests that our simulated disks
which are made up of at least a Neptune mass in debris (those that are massive enough to have measurable
effects on the lifetime of the system) would have planetesimals that are larger than those that are
predicted by the collisional cascade. Therefore, while our planetesimal masses are close to those used
in the `Nice' model, they could be more massive than those in HR 8799's actual disk. This may result in
more stochastic interactions between the planetesimals and planets. However, the planetesimal masses
suggested by the collisional cascade are a distribution which could reasonably involve both smaller or
larger planetesimals then those which we have suggested.

\section{Integrator}
All simulations were run with the software package
\textit{QYMSYM}\footnote{See author's Web site for source code.}. \textit{QYMSYM} is a
GPU-accelerated hybrid second order symplectic integrator which permits close encounters similar
to the \textit{Mercury} software package developed by \citet{chambers99}. Like \textit{Mercury},
\textit{QYMSYM} uses a second order symplectic integrator to advance the positions of all
particles in a simulation in the typical manner described by \citet{duncan98} and
\citet{levison94}. Exploratory work on symplectic maps for N-body integrators was elucidated by
\citet{wisdom91}. Additional analytical work on the formulations of the symplectic integrator
can be found by \citet{saha92} and by \citet{yoshida90,yoshida93}. Unlike these first symplectic
integrators but similar to \textit{Mercury}, \textit{QYMSYM} flags any particle from an integration
when it is deemed that it has a close approach to another massive particle during a time step.
These are then integrated separately with an adaptive step-size conventional integrator. The
criterion for closest approach is decided based on the Hill radii of the interacting objects.
\textit{QYMSYM} uses the $4^{th}$ order Hermite integration scheme detailed in \citet{makino92}
rather than the Bulirsch-Stoer algorithm used in \textit{Mercury}. See \citep{moore11} for more
details on the \textit{QYMSYM} integrator.

We note that due to the nature of our CUDA based code, it is not possible to run less than one
block worth of particles, even in the case where we wish a massless debris disk. Furthermore,
due to memory latencies, low particle counts are not run efficiently on the GPU. Currently, we
do not have the ability to disable the $O(N^2)$ kick computation or a way to offload low particle
counts to the CPU. This is something that could be rectified in future revisions of the code.
Additionally, because of the compilation process, our code is restricted to operating on
machines which have GPU's attached.

These two caveats have an effect on the number and size of the parameter space that we are able
to explore.

The inefficiency of the GPU at low particle count and single block restriction has the impact of
reducing the length for which we can integrate our debris-less test case. We do not want to run
multiple integrators. Two hybrid-symplectic integrators will not get identical answers unless the
collision integration method is the same. Rather than use another hybrid-symplectic integrator
(like \textit{Mercury}) to run debris-less test cases, we continue to use our own code. However,
this does make it difficult to run simulations with as many orbital periods as those possible
with integrators which can be run with very low particle counts.

The GPU requirement also makes it more difficult to test a wide parameter space. A node must
be equipped with a GPU in order to be able to run our code. The GPU clusters that we have
access to are much smaller than the CPU clusters that are available. This limits the number
of possible trials.

Last, we note that our integrator is truly $O(N^2)$ - all particles feel the effects of all
others. This is unlike many integrators available in the field which typically do not compute
planetesimal-planetesimal effects. This makes our simulations more precise, but does
increase the arithmetic intensity.

All simulations were run on either dual or quad core Intel Core2 Duo architecture CPU's with either
GT200 of GF100 architecture NVIDIA GPU's, both of which are CUDA capable double precision
architectures. The code can easily be optimized to run on any Linux kernel 2.6+ distribution with
appropriate GPU hardware.

\subsection{Accuracy, integration parameters and eccentricity corrections}
Timestep sizes were set to 0.016 (out of a possible 2$\pi$ orbit) corresponding to 42 days in
simulations including HR 8799e and 93 days in simulations without. We used smoothing lengths which
correspond to radii at least a few times larger than the size of the planets assuming a planet and
planetesimal density of the order of $1 g/cm^3$ and disregard any simulation during which the energy
conservation (given by $\Delta E/E$) dips below $1.0 \times 10^{-4}$. Additionally, we are only
interested in the time to the onset of first instability, and do not need to concern ourselves
with larger energy errors that may occur after a planet-planet interaction or ejection.

In a few simulations, very high eccentricities were detected for planetesimals which had been
ejected from the system during a close approach. Because our integrator explicitly conserves
angular momentum through the use of f and g functions when solving Kepler's equations (see
\citet{moore11} for more details), these large eccentricities and their corresponding high
velocities produced a drift in the location of the center of momentum. This is most likely
due to the way a hybrid symplectic integrator typically updates the particles positions. If
a particle has its position updated such that it is now in close proximity to another particle,
it would feel a large force. Typically this force is removed by reverse integration of the
Hamiltonian and the particle is then moved to a more accurate integration regime (the Hermite
integrator in our case). However, if by chance that particle is updated to be very close to
another, the assumptions made in the perturbation theory used to split the Hamiltonian into
components, namely that Keplerian motion is dominant over interparticle forces, no longer
applies. This means that the particle will not be correctly reverse integrated. This problem
is common to many symplectic integrators.

Once a planetesimal has been ejected with a very high eccentricity this phenomenon is more
likely to repeat. The high eccentricity means that its angle of impact will be nearly
perpendicular to the motion of the other orbiting objects exterior to its current position.
This newly large velocity relative to the size of the collision detection criteria makes
it more likely for a particle to be updated from outside the collision detection criteria
to be in close proximity to the colliding particle without being previously removed.

Including a larger collision detection criterion or reducing the timestep size of the
simulation could resolve this problem, but at greatly increased simulation time. Decreasing
timestep size goes directly with total simulation time. Due to the high velocities achieved
by the particles, it would need to be decreased significantly. Increasing the collision
detection radius, which is some factor of the Hill radius, will force more particles to be
offloaded to the significantly slower Hermite integration routines. Not only is the Hermite
integrator $O(N^2)$ (and therefore suffers from non-linear increases in simulation time
based on increasing particle count), but the number of particles included in the Hermite
integrator will increase by a factor that goes with the collisional volume - even small
increases in the collision detection criteria can lead to significant increases in particle
counts. Alternatively, increasing the smoothing length can partially alleviate this
problem but has the negative effect of smoothing out many of the important
planetesimal/planetesimal dynamics.

To correct for this occasional error without greatly increasing simulation time, we wrote an
additional check that could remove planetesimals which had eccentricities above an arbitrary
threshold or those that collided with the star. Additionally, particles which are no longer
bound and have a high enough semi-major axis are removed. Energy error checks are common in
comparable literature (see appendix of \citealt{raymond11} for an example) as is planetesimal
removal due to ejection.

This check will have no effect on simulations in which planetesimals are not ejected or ejected
with realistic velocities and eccentricities that are still close enough to interact or collide
with another particle. Only simulations which have particles with extremely large eccentricities
will be quantitatively altered. As we would expect, in simulations where these particles are
removed, it appears that the dynamics are qualitatively identical, only without the aforementioned
drift in center of momentum. We also note that the center of mass is conserved to a high degree
of accuracy and we maintain a satisfactory level of energy conservation throughout either type of
integration. Conservation of center of mass is typically on the order of one part in $10^{12}$ or
better. We therefore present the results of the corrected and uncorrected simulations together -
using uncorrected simulations when no drift is detected in their outputs and corrected versions
elsewhere.

The above timestep choice and numerical energy integration error are comparable to those reported
by \citet{fabrycky10}.

\subsection{Simulation configurations of HR 8799}
We set up two initial configurations to test in our simulations detailed in each
respective section. Generally, given the importance of the 4:2:1 MMR in previous work, our
simulations begin with this planetary configuration and have an additional debris disk or
alternatively use a somewhat modified version of this planetary configuration. Minimally, the
simulations have the innermost two planets in a 2:1 MMR.

Three planet (b,c,d) plus debris disk simulations were based off of the stable configurations
discussed by \citep{fabrycky10}, using planet masses of 10, 10 and 7 $M_{Jup}$ for planets d, c and
b. Four planet (b,c,d,e) plus debris disk simulations are based off the predicted orbital elements
found by \citep{marois10} and use masses of 7, 7, 7 and 5 $M_{Jup}$ for planets e, d, c and b. Our
debris disk is simply a uniform distribution of 1024 equal mass particles. Depending on total disk
mass desired, the planetesimal mass was modified accordingly.

This later set of initial positions and masses reflects the most current observational results
by \citet{marois10} by including the fourth planet and the corresponding reduced masses required
from the dynamical simulations that were run. The three planet configuration, while no longer
representative of those more recent observations and simulations, is both practical and
illuminating for several reasons.

First, we note that the planet which is not included in the three-planet configuration is the
most recently discovered innermost object. Unless its presence destabilizes the system, we expect
its effect on the migration rate of the outermost planet to be small in those simulations due to
the proximity and mass of the debris disk to the outermost planet. Additionally, we only use the
three planet configuration for simulations in which the system is arbitrarily placed in an unstable
configuration from the beginning. The four planet configuration has much smaller regions of
stability. While the addition of the fourth planet would make it even easier for us to create an
initially unstable configuration which shares orbital elements similar to those that are observed,
it makes it significantly more unlikely for a system to move from an unstable region to a stable
region via orbital migration. Last, we recall that the recent work by \citep{soummer11} which
reduced Hubble Space Telescope observations suggests best orbital fits for HR 8799b,c,d that
coincide with the 1:2:4 mean motion resonance. This data agrees with previous work that fits
only the three planets, but does not agree with dynamical simulations which included all four
planets \citep{marois10}. \citealt{marois10} found that the innermost three planets, HR 8799c,d,e
were the most likely planets to share mean motion resonances with b excluded. There is as of yet
no consensus on the possible orbital elements or dynamical structure of HR 8799 other than the
most long-lived systems having the innermost three planets in a 4:2:1 mean motion resonance or
at least having the innermost two planets in a 2:1 mean motion resonance. Therefore we assume
that the 2:1 mean motion is the primary initial requirement for our simulated systems.

Because we are measuring the effects of planetary migration on system stability to determine if
all unstable configurations remain so, it is useful to run the three planet system - it has larger
regions of stability for the planets to migrate into. Due to the computational intensity of the
simulations and their corresponding time requirements, we were only able to run on the order of
hundreds of simulations over a few months rather than tens or hundreds of thousands. Given this
limitation, it proved difficult to migrate large numbers of four planet configurations into regions
of stability. This lower simulation number and corresponding low number of stabilized configurations
would introduce a fine tuning problem to our analysis so we do not discuss those in greater detail,
instead focusing on the three planet configurations for the stabilization through migration
scenarios. This fine tuning issue could potentially be resolved via more stable initial conditions
which would require significantly less planet migration to move from regions of instability to
regions of stability. This issue is discussed in further detail in our results section.

In the three planet configurations the inner disk edge has a semi-major axis of $a_{min}$ = 2.5
and an outer disk edge of $a_{max}$ = 7.0, corresponding to separations of 60 and 170 AU.
In the four planet configurations the inner disk edge has a semi-major axis of $a_{min}$ = 6.14
and an outer disk edge of $a_{max}$ = 20.69, corresponding to separations of 90 and 300 AU.
This second set matches the estimated outer debris disk's observed inner and outer edges.

The inner disk edge with reduced semi-major axis for the three planet configuration was used
to encourage rapid planet-planetesimal crossings and therefore rapid migration. As mentioned
previously, we forced our three planet configuration to become unstable on very short
timescales. Because of this, we require rapid migration in order for any outcome other than
planet-planet scattering to be a possibility. The inclusion of a debris disk at a position
more coincident with that which is observed would be possible with either faster or more GPUs.
This is because configurations that are stable on longer timescales - which allow for reduced
amounts of planet migration to be necessary in order to stabilize an unstable configuration -
could be used. We could additionally begin to experiment with the inclusion of the fourth
planet as well as keep the outermost planet at its observed position rather than moving it
in to encourage the system to become unstable. We found both the removal of the innermost
planet and movement of the inner disk edge were helpful in finding these post-migration stable
configurations.

Other simulation parameters include the Hill factor for encounter detection and the K function
(see \citet{moore11}), both of which are set to 2.0. The K function is an arbitrary function
which weights certain part of the Hamiltonian to be integrated in order to allow it to be
broken up into evolutions operators which are otherwise not possible. When the distance between
two particles is large, the value of K goes to zero while when the distance between two particles
is small, the value of K goes to 1 (or vice versa). When K or (1-K) are multiplied into the
respective separated Hamiltonians described in \citep{moore11}, it prevents either from becoming too
large and breaking the perturbation theory used to separate them. K is a function of Hill radii,
but was created by trial and error.

The distribution of planetesimal semi-major axes is flat with probability independent of a within
$a_{min}$ and $a_{max}$. The initial eccentricity and inclination distributions were chosen using
Rayleigh distributions with the mean eccentricity $\overline{e}$ equivalent to twice the mean value
of the inclination $\overline{i}$ and $\overline{i}$ = 0.01. The initial orbital angles (mean
anomalies, longitudes of pericenter and longitudes of the ascending node) were randomly chosen.

A group of 150 simulations with three planets were run, 58 using the eccentricity correction while
92 without. Disk mass was varied from $1.0 \times 10^{-30}$ to $10$ $M_{Jup}$, although we only
present results from a disk mass of $1.0 \times 10^{-3} M_{Jup}$ and larger here. 18
simulations with four planets were run, half using the eccentricity-correction while the other half
without. These simulations differed only in the initial conditions of the planetesimals.

\section{Simulations of initially unstable planetary orbital architectures}
Does the addition of a relatively massive planetesimal disk allow for an increase in stability
timescale? To answer this question we created an unstable initial configuration for both
the three and four planet configurations of HR 8799. In either case, this was done by starting with
actual observed orbital elements found by \citet{fabrycky10} and \citet{marois10} and reducing the
outermost planet's semi-major axis in small increments until the system was unstable on the time
scale of thousands of orbits or less. In the three planet simulations, this gives an initial
configuration with all three planets having orbital elements the same as those reported by
\citet{fabrycky10} except that the outermost planet's semi-major axis was reduced by $14\%$. For
the four planet simulations, the planets' orbital configuration was identical to those found by
\citet{marois10} but the outermost planet has a semi-major axis reduced by $8\%$. This arbitrary
and largely unstable configuration was chosen to both reduce simulation time as well as allow for
pronounced effects by migration. Due to the previously mentioned limitation in the number of
total simulations it is possible to run and the extremely limited regions of stability available
to the four planet configuration, we opted to run far more three planet simulations than four
planet systems and present only that data.

Simulations both with and without eccentricity corrections are presented simultaneously and
see no significant difference between them.

\subsection{Results}
\begin{figure*}
\includegraphics[width=150mm]{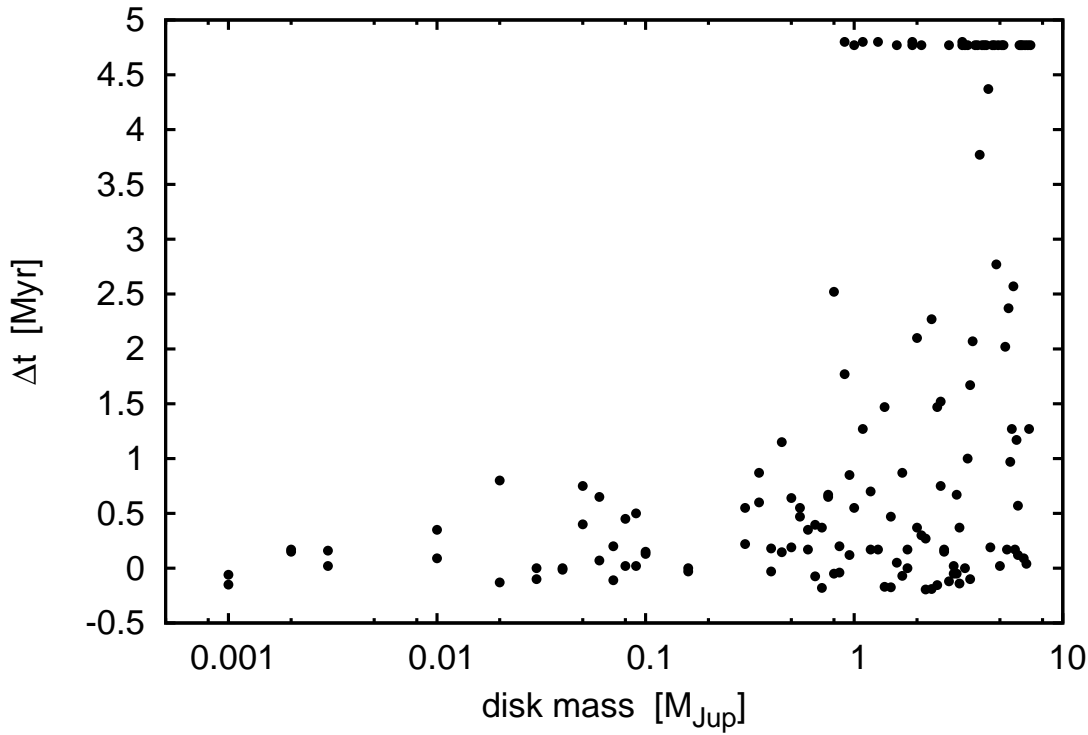}
\caption{This figure shows the difference in time until a planet-planet encounter for a group of
HR 8799-like simulations with a debris disk from an identical simulation with no disk mass versus
the amount of mass in that particular disk. An intentionally unstable solution was used in order
to quickly show the effects of a planetesimal debris disk. All simulations use orbital elements
taken from \citet{fabrycky10} for the inner two planets (which are in a 2d:1c resonance). The
outer planet, HR 8799b, has a semi-major axis which is reduced by 14$\%$ from approximately 68 to
58.6 AU in order to achieve a planet-planet close approach within a couple hundred thousand years.
$\Delta$t represents the increase or decrease in time before the first planet-planet encounter.
Simulations were halted after the first planet-planet encounter or after 5 Myr.}
\label{fig:stabilizing}
\end{figure*}

In figure \ref{fig:stabilizing} we plot the difference between the stability time for each simulation
and that of a system lacking a debris disk. The stability time is the time to first planet-planet
encounter. The time difference is plotted as a function of disk mass. Simulations are run for a maximum
of 5 Myrs. If no encounters had taken place in that time, they are plotted as upper limits in the figure.

\begin{figure}
\includegraphics[width=85mm]{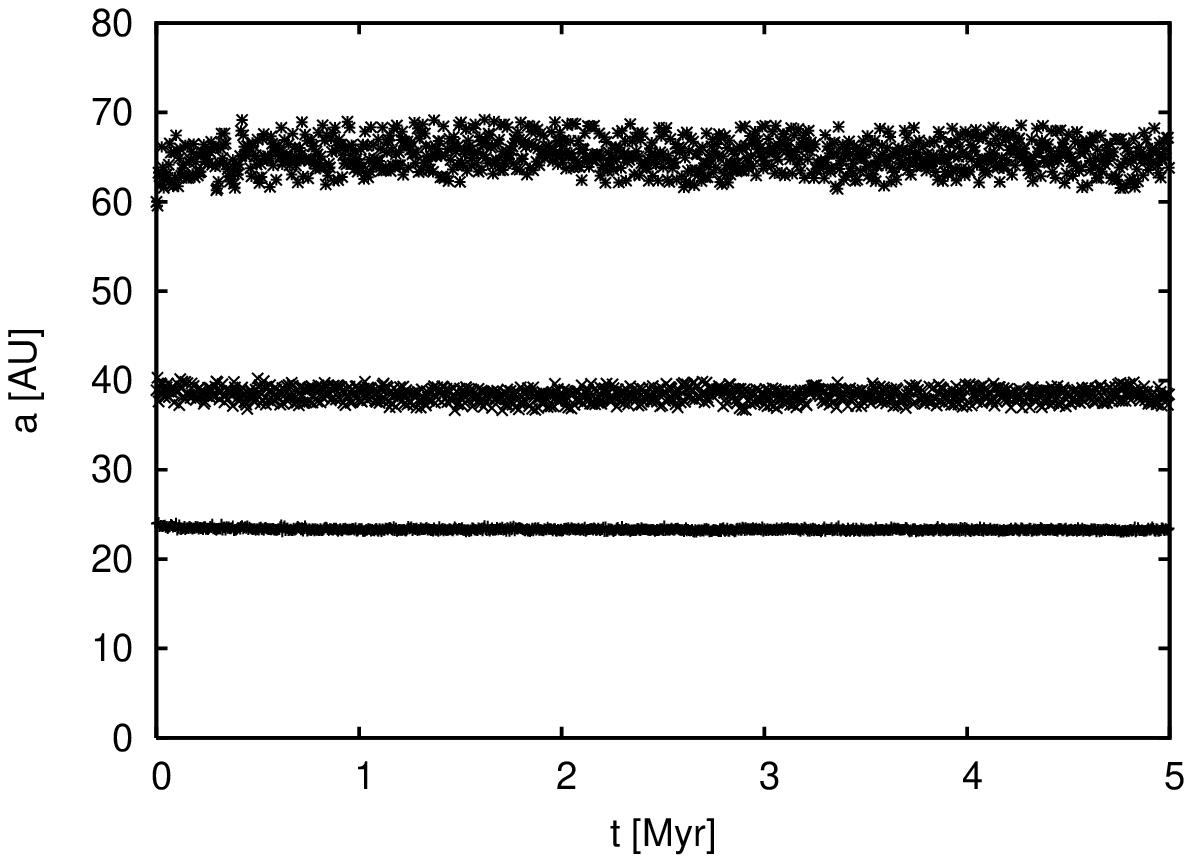}
\includegraphics[width=85mm]{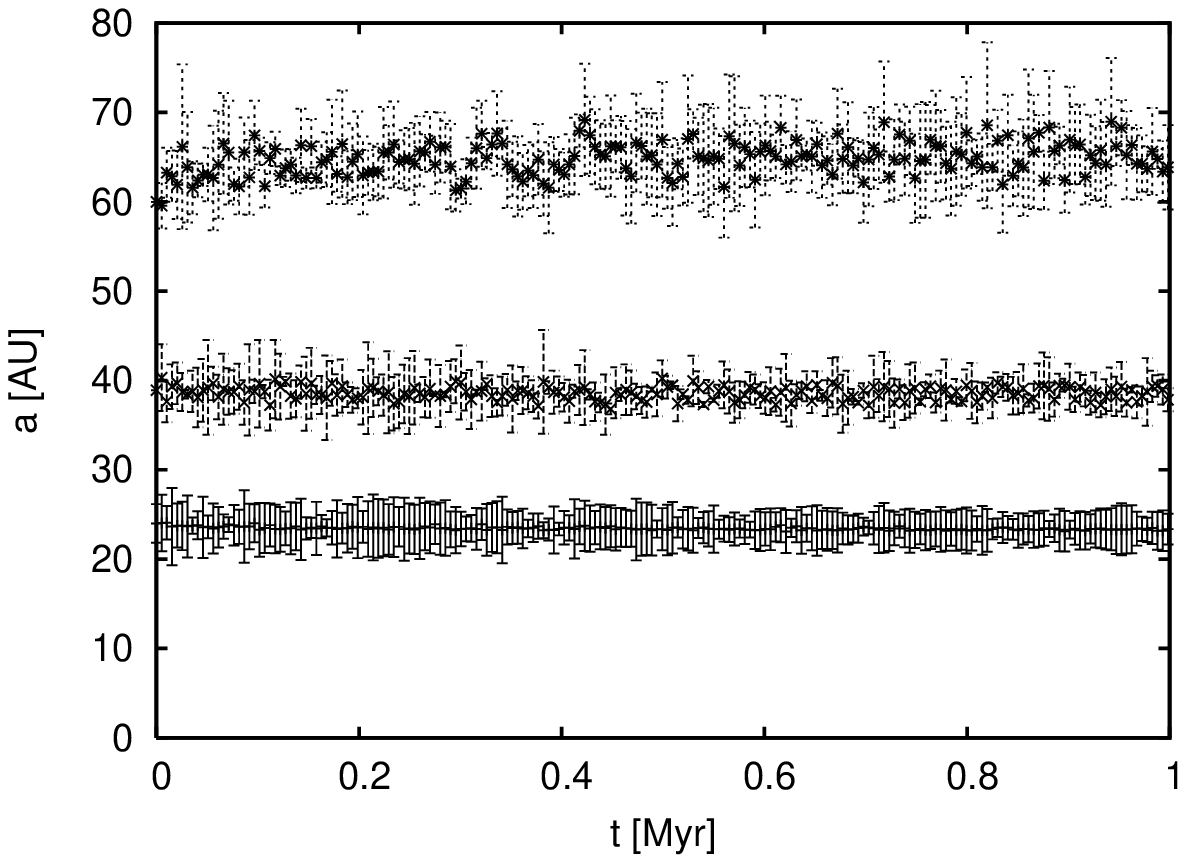}
\caption{In this figure we show the evolution in semi-major axis of an example three planet configuration
over time. a) Shows the evolution in time of the semi-major axis of the three planets over the entire 5 Myr
simulation. b) Shows the evolution in time of the semi-major axis of the three planets over the first Myr.
In b) we also include error bars that indicate the maximum and minimum separation (given by a(1+e) and
a(1-e) respectively) of the planets given their eccentricity. This second image more clearly shows the
migration of the outermost planet planet from its initial position at 58.6 AU to approximately 65 AU over
the first few hundred thousand years. In this example the total disk mass was $1.6 M_{Jup}$ and the
configuration remained stable over the entire 5 Myr simulation.}
\label{fig:3p-migration}
\end{figure}

In figure \ref{fig:3p-migration} we plot the evolution in time of the semi-major axis of the three
planets for a sample simulation. In this case the total disk mass simulated was $1.6 M_{Jup}$.

Figure \ref{fig:stabilizing} shows that at low disk masses the difference in lifetime is near 0. Only
massive disks with masses of ten percent or more of the outermost planet can remain stable until the end of
the simulation. Given enough mass, migration via planetesimal scattering can take a tightly packed
configuration of HR8799 b, c, d and e, or in the case of figures \ref{fig:stabilizing} and
\ref{fig:3p-migration}, simply b, c and d, and migrate the outermost planet into a more stable
configuration. The total migration for the outermost planet in system configurations which remained
stable over the 5 Myr simulation ranged from 5 to 15 AU in semi-major axis depending on total disk mass.
This migration occurred over a few hundred thousand years.

If stable regions are small compared to unstable regions, migration is unlikely to put an initially
unstable configuration into a stable region. Only large migrations could significantly increase the
stability time. This is consistent with what we see; rapid migration caused by a massive disk is able
to significantly affect the lifetime of our simulations. Migration is expected to reduce planetary
eccentricities and this could increase stability \citep{lissauer93,fernandez96}. We searched for signs
that the outermost planet's eccentricity decreased during migration. In figure \ref{fig:3p-migration}
we see oscillations in eccentricity of the outermost planet were large throughout the simulation due to
the proximity and high masses of the inner planets. We therefore attribute the increased stability time
to the wider separation rather than any eccentricity damping.

As a planet migrates outwards it can cross mean motion resonances with other planets. For simulations with
short lifetimes, we searched for evidence that resonant crossing caused instability. We examined any
relevant $1^{st}$, $2^{nd}$, $3^{rd}$ and $4^{th}$ order resonance possible between the outer planet and
innermost planet as well as the outer planet and the middle planet. We saw no large semi-major axis or
eccentricity jumps caused by resonant crossing so they are unlikely to be the cause of instability in the
simulations with shorter lifetimes. Resonances near the outer planet's position are primarily $2^{nd}$
or $3^{rd}$ order and cause smaller eccentricity jumps due to the weaker dependence on mass
\citep{quillen06}.

As discussed in our section on simulation parameters, our choice of unstable configurations was done
in part to reduce the amount of simulation time. With an initial planetary configuration so far from a
region of stability, a large amount of migration is required in order to move the planet to a more stable
region of phase space. Because of this, only a massive debris disk can migrate the outermost planet
sufficiently far to increase the lifetime. 

Another concern with this set of simulations is that with a planetesimal count of 1024 and a total disk
mass of at least a Jupiter mass, interactions between planetesimals and planets would cause stochastic
migration \citep{zhou02}. We would have expected such large planetesimal masses to cause instability
rather than increase the lifetime of the system. Therefore, we have no reason to suspect that stochastic
behavior is a problem. Because the typical disk mass required to stabilize the system is unrealistically
large there may be no need to resolve the disk more accurately. As mentioned previously, an approximate
upper limit for debris mass is thought to be on the order of $10^{-3} M_{\odot}$, or approximately one
Jupiter mass. For HR 8799, this is effectively the minimum amount of mass required for any observable
effects in our simulations.

\section{Simulations of initially stable planetary orbital architectures}
In this second set of simulations we examine the impact of a planetesimal debris disk on the
stability time of a configuration of planets which was stable for about 150k orbits (or around
7 Myr) sans debris disk. In these simulations we begin with all four planets situated in an
orbital configuration based off the current estimated positions by \citet{marois10}. This four
planet configuration is generally extremely unstable unless situated in a 4e:2d:1c MMR. As before,
smoothing lengths are set to sizes a few times the radii of the planet to prevent the unrealistic
forces possible if two particles form a very tight binary or collide within what would have been
considered an approximate planetary radius. Energy error is typically even lower in these simulations
with a $\Delta E / E$ at $1.0 \times 10^{-5}$ or $1.0 \times 10^{-6}$ due to reduced planetesimal masses.
We stopped simulations after their first planet-planet encounter regardless of the eventual fate
of the interacting planets. Stability time in the histogram is the time to first planet-planet
encounter.

\begin{figure}
\includegraphics[width=85mm]{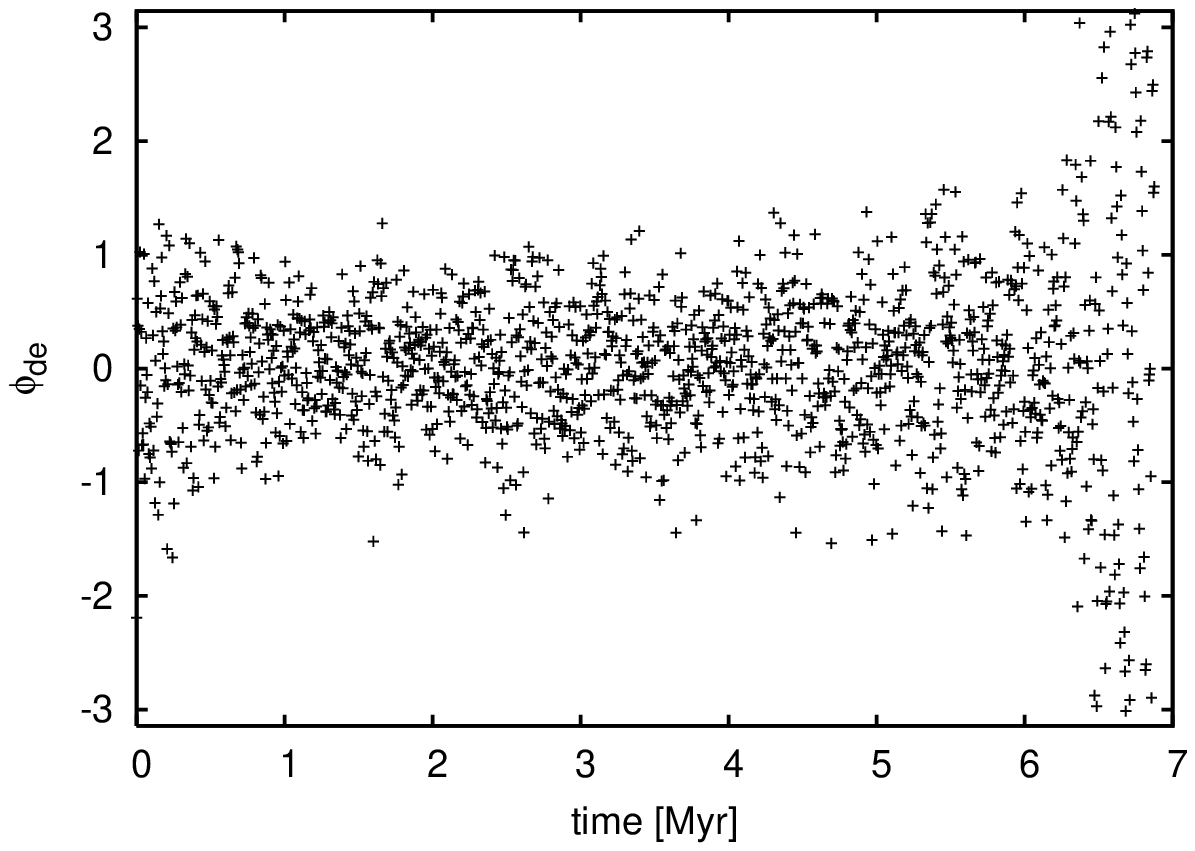}
\includegraphics[width=85mm]{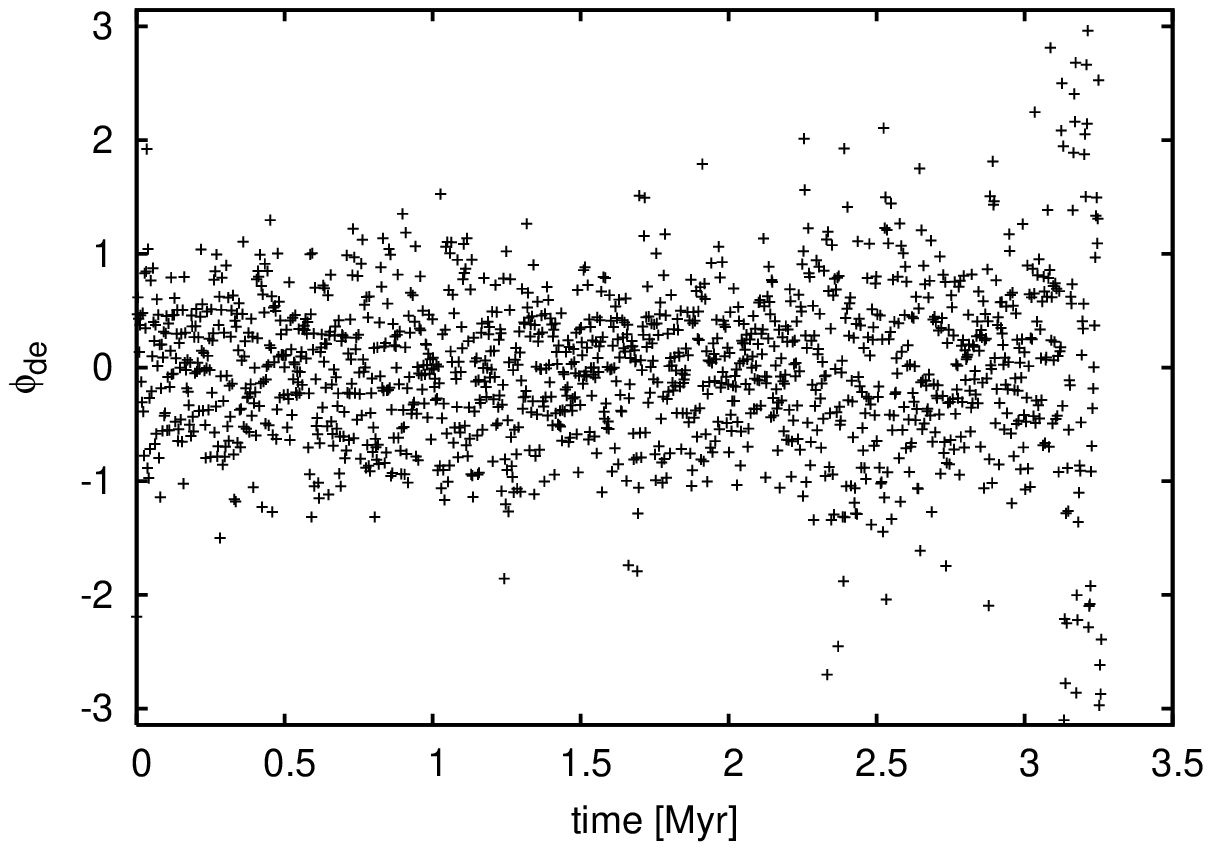}
\caption{Here we show the resonant angle for 2:1 mean motion resonance for HR 8799d
and e as function of time.
a) For a simulation without a debris disk.
b) For a simulation with a debris disk that has a mass of 1/100th the outer planet.
The initial conditions for the planets were stable for 7 Myrs. In b) the extreme values of
$\phi$ slowly begin to increase over the length of the simulation. This effect is not present in a).
The resonant island is so small that even a disk mass of $1\%$ the outer planet is capable of
effecting the systems dynamics. The resonant angle plotted is given by
$\phi _e = 2 \lambda _d - \lambda _e - \varpi _e$.}
\label{fig:angles}
\end{figure}

We began by simulating the four planets with an initial orbital configuration comparable to that
by \citet{marois10} with massless debris and plotted the resonant arguments. We see constrained
resonant angles which indicate the presence of a 2:1 MMR between the inner two planets, shown in
figure \ref{fig:angles}. While \citet{marois10} did not directly show the resonant argument to
illustrate that HR 8799d and e were in resonance, our plot is similar to Figure 10 by
\citet{fabrycky10} although for HR 8799e and d rather than d and c. The integrator confirms that
this planetary configuration is in resonance as suggested by \citet{marois10}.

Matching the exact age found by \citet{marois10} is difficult despite the fact that there is no
longer a requirement of planetesimals with mass. This is due to the way the integrator operates.
Specifically, it is not possible to integrate less than a certain number of particles, depending
on compile time and hardware restrictions. For more details on the exact nature of the code,
we refer the reader to \citet{moore11}. However, while the control simulation is not as long
lived as those found by \citet{marois10}, it is still stable on long time scales ($10^5$ orbits).

This simulation is then compared to 18 simulations with Neptune mass disks. These 18 simulations are
identical in terms of planetesimal mass, total disk mass, initial outer and inner disk edge, and initial
planetary orbital configuration. However, the planetesimals initial orbital configurations have been
randomized 18 different ways.

\begin{figure}
\includegraphics[width=85mm]{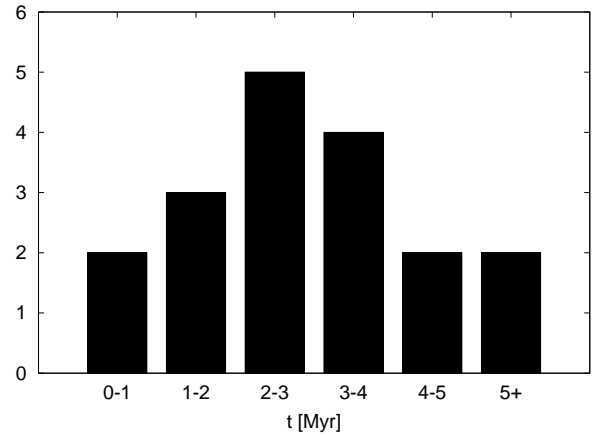}
\caption{Here we show a histogram of time to the onset of instability, which we define as planet
interaction or ejection. The disk mass is about one Neptune in mass. The planets are initially in
a configuration that is stable for approximately 7 Myrs. We see that the distribution of lifetimes
is very broad and many of the simulations had lifetimes less than this. The reduction in lifetimes
was as large as $\sim6$ Myrs.}
\label{fig:histogram}
\end{figure}

\begin{figure}
\includegraphics[width=85mm]{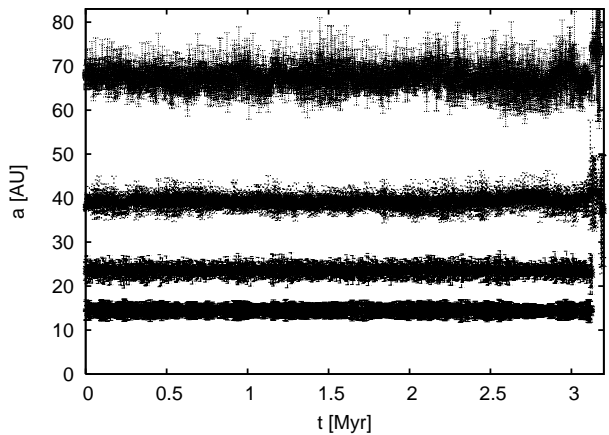}
\caption{In this figure we show the evolution in time of the semi-major axis of an example 4
planet configuration. This particular simulation became unstable in a little over 3 Myr. We include
error bars that indicate the maximum and minimum separation (given by a(1+e) and a(1-e) respectively)
of the planets given their eccentricity. In all four-planet simulations the total disk mass was equivalent
to Neptune.}
\label{fig:4p}
\end{figure}

The histogram shown in figure \ref{fig:histogram} shows the distribution in lifetimes of the
eighteen simulations each with a Neptune mass disk. Figure \ref{fig:4p} shows the evolution in
time of the semi-major axis of all four planets of a sample simulation. The error bars indicate
the minimum and maximum separations.

\subsection{Results}
In figure \ref{fig:histogram} we see that in 16 of 18 simulations the stability time has decreased by at
least 2 Myrs from the initial `stable' configuration which had its first planet interactions after
7 Myrs. We see a broad distribution of lifetimes. Even though a Neptune mass disk is only about
$1\%$ the mass of the outer most planet it has an effect on the stability time. 

With a Neptune mass disk each simulated planetesimal has a mass of about 8 Plutos. Debris disk models
estimate the top of the collisional cascade to contain similar (but slightly smaller) size bodies. For
example, \citet{wyatt02} argued for a distribution of planetesimals in the Fomalhaut system with a
range of sizes between 4 and 1000km. As discussed previously, this second value roughly coincides with
Pluto's diameter within a  factor of a few of what we used in our simulations. Stochastic perturbations
are expected in real systems; however they will be somewhat smaller than those present in our simulation
do to our factor of 8 difference in mass of planetesimals.

Figure 3 by \citet{gozdziewski09} demonstrates why a planetesimal disk with a total mass around that of
Neptune can have such a pronounced effect on the stability time. In this plot, dark regions indicate
highly likely configurations of eccentricity and semi-major axis for planet d, whereas yellow regions
indicate strongly chaotic systems. Here we see that semi-major axis changes of less than 0.1 AU can move
planet d from a stable region to a strongly chaotic one. Similarly, minor changes in eccentricity can
also have a large effect. \citet{fabrycky10}'s Figure 7 shows a similar behavior. Note that a change in
current separation of planet d by 0.05 in the figure (corresponding to semi-major axis change of 1.22 AU)
can decrease the stability time by as much as four orders of magnitude. A three order of magnitude change
in stability time is possible with a semi-major axis increase of approximately half an AU. In either
example the regions of stability have sharp edges between stable and chaotic solutions.

We would expect a disk mass of $1\%$ the mass of a planet to be able to migrate a planet roughly $1\%$
of the semi-major axis if a majority of its angular momentum is transferred to the planet via scattering.
If this were the case than a Neptune mass disk would be able to migrate the outer planet (which has a mass 
of $5 M_{Jup}$ at 68 AU) approximately 0.5 AU. This is more than enough to vary the lifetime by many 
orders of magnitude. Therefore we looked for the number of planetesimals which had become orbit crossing
by the time each simulation had become unstable. In all 16 simulations, at least $15\%$ of the disk
mass was orbit crossing. This corresponds to a minimum of 2.5 $M_{\oplus}$. By starting a simulation near
the edges of a region of stability, we see that even lower mass debris disks can affect the stability time.

\subsection{Pulling HR 8799's planets out of resonance}
We continuously monitor the planets' orbital elements throughout all simulations to see if they
are in resonance with one another. We do this by noting whether or not the resonant angle
librates around 0 or $\pi$. We looked for
$1^{st}$, $2^{nd}$, $3^{rd}$ and $4^{th}$ order mean motion resonances that could exist near
each of the planets' semi-major axis at varying times. We did not observe any strong two
body resonances beyond the 2:1 MMR between e and d, a much weaker 2:1 MMR between d and c, and
no clear resonant angles for b. We attribute this to the overwhelmingly large gravitational
perturbations caused by the massive planets as well as the fact that most of the nearby
resonances for the planets were of higher order. This prevented behavior similar to the `Nice'
model during which migration causes resonance crossing and corresponding eccentricity increases.
Finally, we see that in all simulations when the planets e and d are pulled out of the 2:1 MMR
the system rapidly becomes unstable.

In figure \ref{fig:angles}a we see that the resonant angle for the 2:1 MMR between HR 8799e and
d over the 7 Myr simulation remains largely unaffected until the end of the simulation. However,
in figure \ref{fig:angles}b we note that the extreme values of $\phi$ begin to increase over time,
leading to e and d moving out of resonance and the rapid disruption of the system. In this
particular example, the system's lifetime is decreased by about 50$\%$. It is possible the debris
disk is responsible for pulling HR 8799 out of resonance and so, reducing its lifetime.

\section{Conclusion}
In this paper we discuss how a planetesimal debris disk can effect the stability of the multiple planet
system HR 8799. Two questions are considered; whether it is possible to destabilize a stable planetary
configuration with a planetesimal debris disk and conversely, whether it is possible to stabilize an
unstable configuration with a planetesimal debris disk. We examine both three planet (b,c,d) and four
planet configurations (b,c,d,e).

In three planet configurations which were unstable on short timescales without a debris disk, only massive
debris disks ($10\%$ the mass of the outermost planet) could cause increases in stability time. In four
planet configurations which were stable over long timescales without a debris disk, debris disks of only a
Neptune in mass ($1\%$ the outer planet) can cause large decreases in lifetimes. We attribute this
sensitivity to the small size of HR 8799's resonant region of stability.

While the amount of disk mass required for system stabilization in our simulations is unrealistic, it
is only required to be this large due to the initial conditions. In order to run a large number of
simulations, a highly unstable configuration that rapidly had planetary encounters was required.
Given an initial configuration which is just outside a region of stability, it would be possible to
cause sufficient migration to allow an unstable configuration to become stable with a more reasonable
disk mass.

Similarly, while the disk properties required for system disruption in our simulations are reasonable in
both total mass and planetesimal size, the mass distribution for the planetesimals is not a true mass
distribution. In order to keep total particle number down we restrict all planetesimals to have an
identical 8 Pluto sized mass. We do not believe that this had a large effect on the distribution of
stability times of the system when a disk is included because a realistic mass distribution could
encourage even more stochastic migration if there are even a few planetesimals of greater mass.

These results suggest that HR 8799 may be destined for eventual planet scattering. The very high masses
of the planets causes regions of stability to be relatively small and makes it possible for low mass
debris disks to pull the system out of its mean motion resonances and induce instability.
We also see that migration is much more likely to make the system unstable than migrate the system to a
region of stability. It may be the case that the debris disk has already been responsible for removing
the system from a maximally stable region and is currently near the boundary of instability. All of
these possibilities tend to induce instability rather than stability and suggest that HR 8799 may be
headed towards instability, possibly sooner than would otherwise be predicted.

In this study we have used initial conditions consistent with observed positions of the planets when 
attempting to determine whether a stable orbital configuration could be made unstable by a debris disk.
As we have shown here, a low mass disk can cause evolution of the planetary configuration.
Consequently, in the past the planets could have been in a different configuration. By exploring
different initial conditions future investigations could explore scenarios for the past evolution of
HR 8799 that would be consistent with its current configuration. For example, if the planetesimal
disk is causing the system to move out of resonance, in the past it may have been in a more stable
region of this resonance, instead of on its boundary. Increasing simulation length and particle count
could also increase the resolution of our simulations.

Alternatively, it appears that similar simulations could be used to put upper limits on planetesimal
debris disk mass if it is assumed that the debris disk has a negligible dynamical effect. In the case
of our simulations of HR 8799, planetesimal disk masses would need to be much smaller than one Neptune
in mass.

The stability timescale for an HR 8799-like system is very sensitive to small alterations in planetary
orbital elements. This sensitivity implies that the effects of planetesimal debris disks may not be
negligible for systems with such small regions of stability. However, dynamical simulations of HR 8799
sans planetesimal disk were used to determine the resonant structure of the system and constrain
observations while not including all the required dynamics. The assumption of long term maximum
stability may be erroneous when debris disks could allow for migration of planets from currently
observed stable regions to unstable configurations or vice versa. In general, it would appear that
using dynamical models which do not include the effects of planetesimals to constrain observations may
be unwise when a system is known to have such limited regions of stability. 

Additionally, the long term stability of systems with lower mass disks could still be important. In
these systems the planets will typically be smaller and the regions of stability would be larger.
However, reduced planet mass would increase the migration rate.

{}

\end{document}